# Time-resolved imaging of pulse-induced magnetization reversal with a microwave assist field

Siddharth Rao,[1] Jan Rhensius,[1] Andre Bisig,[2, 3] Mohamad-Assaad Mawass,[2, 3] Markus Weigand,[3]
Mathias Kläui,[2] Charanjit S. Bhatia[1] & Hyunsoo Yang[1]

[1]*Department of Electrical and Computer Engineering, National University of Singapore, 4*
*Engineering Drive 3, Singapore 117576, Singapore*
[2]*Institute of Physics, University of Mainz, Staudinger Weg 7, 55128 Mainz, Germany*
[3]*Max-Planck-Insitut fur Intelligente Systeme, Heisenbergstr. 3, 70569 Stuttgart, Germany*

**The reversal of the magnetization under the influence of a field pulse has been previously
predicted to be an incoherent process with several competing phenomena such as domain
wall relaxation, spin wave-mediated instability regions, and vortex-core mediated reversal
dynamics. However, there has been no study on the direct observation of the switching
process with the aid of a microwave signal input. We report a time-resolved imaging study
of magnetization reversal in patterned magnetic structures under the influence of a field
pulse with microwave assistance. The microwave frequency is varied to demonstrate the
effect of resonant microwave-assisted switching. We observe that the switching process is
dominated by spin wave dynamics generated as a result of magnetic instabilities in the
structures, and identify the frequencies that are most dominant in magnetization reversal.**

Correspondence and requests for materials should be addressed to H.Y. (eleyang@nus.edu.sg)

Magnetization reversal dynamics in patterned magnetic structures have been intensely studied in recent years due to their relevance in high speed switching applications,[1-3] and due to fundamental interests.[4-9] Long-range magnetostatic interactions due to shape effects become stronger as structural dimensions approach the sub-micron scale, giving rise to a rich composition of stationary and propagating magnetization dynamics, comprised of the fundamental and higher order precessional modes. The influence of competing forces such as the anisotropy, exchange, and demagnetizing fields is evident from the creation of magnetic instability regions that precede the reversal process,[10,11] and are well supported by micromagnetic simulations.[12,13] Using magneto-optical techniques, experiments have shown that one way to speed up the switching process is to suppress the precessional motion, by tailoring the field pulses.[14] Time-resolved studies of pulse-induced switching in tapered sub-micrometer elements confirmed the observation of instability regions,[10] and interestingly switched at a lower magnetic field as compared to rectangular structures or structures with flux-closure domains.[15,16] However, the energy required in magnetization reversal by the use of static square current pulses makes it unsuitable to be considered for future applications in magnetization reversal devices.

To overcome this drawback, the idea of microwave assisted switching (MAS) has been proposed and researched widely over the past decade.[17-22] In MAS, a high frequency microwave current pulse is superimposed on the square current pulse, thereby driving the precessional switching dynamics at a higher frequency. Kerr microscopy studies,[12,23] in conjunction with micromagnetic simulations, of microwave assisted switching in large Py ellipsoids suggest that the reversal is initiated by the generation of inhomogeneous spin wave modes in confined structures leading to domain nucleation, and magnetization reversal through domain wall propagation. MAS has also been explored by electrical techniques through ferromagnetic resonance (FMR) measurements,[17-19] or by using a microbridge superconducting interference device (SQUID).[9] The scanning transmission X-ray microscopy (STXM) technique is an imaging technique that possesses the necessary spatio-temporal resolution to identify the physics behind the MAS process.

In this work, we present direct time-resolved images of the microwave assisted magnetization reversal process in a micron-sized Py ($Ni_{78}Fe_{22}$) element using STXM. The element is excited by a microwave signal superimposed on a pulse field due to a square current pulse. By varying the microwave excitation frequency, we identify the natural resonance frequency region of the magnetic element in which the switching is most efficient. Using micromagnetic simulations and subsequent analysis, we show that domains are formed around the foci of the elliptical element due to spin waves and couple with FMR frequencies at the edges of the element giving rise to further spin waves of edge modes and fundamental precession modes.

**Results**

**Direct dynamic imaging of MAS in patterned Py elements**. The elements studied in this experiment are elliptical structures patterned on a film stack of Ta (2 nm)/$Ni_{78}Fe_{22}$ (Py, 30 nm). The elements were patterned in two different sizes such as a $4 \times 0.4$ $\mu m^2$ element aligned at 45° to the current-carrying stripline (Figs. 1 and 2), and a $6 \times 0.7$ $\mu m^2$ element aligned at 30° to the current-carrying stripline (Fig. 3). The ferromagnetic elements were fabricated beneath a 4 $\mu m$ wide Ta (5 nm)/Cu (150 nm) stripline on a 75 nm thick silicon nitride ($SiN_x$) membrane. The Py elements were patterned by standard electron-beam lithography and lift-off processing (see Methods). The current excitations in the stripline consist of a unipolar square pulse of 2 ns duration superimposed on a 4 ns long, 1.8 GHz sinusoidal microwave input pulse with a 1 ns offset in time, as seen in Fig. 1(b). The Oersted fields generated due to both the unipolar square pulse and the superimposed microwave signal are along the $y$-axis, therefore the initial state destabilizing magnetization dynamics is observed primarily along the $y$-axis. The pulse field applied during this process was -75 Oe and the microwave field was 30 Oe peak-to-peak. The microwave frequency was varied among four values, 1.5, 1.8, 2, and 2.5 GHz for both elements. At these fields, microwave-assisted switching was only successful at 1.8 GHz in the $4 \times 0.4$ $\mu m^2$ element, but was less critical in the $6 \times 0.7$ $\mu m^2$ element. The excitation frequency dependence of the switching process is discussed in greater detail later. The in-plane component of the magnetization was dynamically imaged employing time-resolved STXM by pump-and-probe technique (see Methods). Magnetic contrast is obtained by taking advantage of the X-ray magnetic circular dichroism (XMCD).[24]

The black and white contrast in the inset of Fig. 1(b) corresponds to the direction of the magnetization along the $x$-axis of the element. Figures 1(c-h) show the time evolution of a typical reversal process in the element measured experimentally with both the square pulse and microwave fields. In the experiment, $t = 0$ is defined as the time of the onset of the 4 ns-long microwave pulse. The square pulse is turned on at $t = 1$ ns and reaches its maximum value at $t = 1.3$ ns. The irregular shape of the square pulse seen in Fig. 1(b) is due to imperfect reflection of the signals in the electrical circuitry. Figure 1(c) indicates the state of the element before switching at $t = 1.27$ ns, followed by domain nucleation at the foci of the elliptical element at $t = 1.44$ ns seen in Fig. 1(d). The formation and taper angle of the domains is influenced by both the shape-dependent demagnetizing field in the element and the consequent instability regions formed as a result of spin wave creation and propagation.[10,11] These domains gradually undergo relaxation and expansion processes as seen at $t = 1.77$ ns in Fig. 1(e). The sequence of the domain expansion processes seen in Figs. 1(e-g) indicates that the element does not follow a uniform, coherent precession switching model. The edges of the element remain unswitched until $t = 2.43$ ns (Fig. 1(g)) and reverse only at the end of the square pulse. This is attributed to the significantly higher demagnetizing field at the edges of the element (see Supplementary Information Fig. S1). Even though micromagnetic simulations cannot include magnetic and surface defects that contribute to domain pinning effects and other non-uniformities in the switching sequence, we have performed simulations using the object oriented micromagnetic framework (OOMMF) to confirm a reversal process that is largely similar to that of experiment (Figs. 4(a-f); see Methods for further details). The applied field direction in simulations is the same as that of experiments as shown in the inset of Fig. 1(b). The simulation results will be discussed in further detail later.

**Influence of microwave assist fields in magnetization reversal**. In order to demonstrate that the assistance of the microwave pulse is significant in switching the element, a set of control experiments were performed. In the first control experiment, only the square pulse excitation was applied to the stripline without the microwave excitation. The current pulse flows along the $x$-axis (from left to right), and the Oersted field is generated along the $-y$-axis. Figures 2(a-f) show the time-resolved images of the magnetic element at different time instants after the application of the square pulse at $t = 1$ ns. Clearly, the element undergoes partial switching along the foci of the element. However, once the pulse excitation is removed, the domains gradually annihilate and the element relaxes back to its initial state. In the second control experiment, only the microwave excitation was applied to the stripline and we see that a very weak excitation is generated in the element as shown in Fig. 2(g-i). A full switching was not possible in this case, either. These control experiments confirm that a full switching is only possible with the help of a microwave pulse on top of a square pulse field.

**Influence of microwave excitation frequency on magnetization reversal process**. The control experiments established the necessary contribution of a microwave field in magnetization reversal studies. However, the frequency of microwave excitation also determines the probability of successful switching. To investigate further, the $6 \times 0.7$ $\mu m^2$ elements were subjected to microwave current pulses of four different frequencies varying from $1.5 - 2.5$ GHz. Time-resolved images of the switching attempts at these frequencies are shown in Fig. 3(a). Complete magnetization reversal is observed at frequencies $f = 1.8$ GHz, 2.0 GHz, and 2.5 GHz through different switching evolution processes, however, at $f = 1.5$ GHz partial magnetization reversal of the element is followed by relaxation to the initial state. It is well-known that the absorption of microwave energy is most efficient at the natural resonance frequency of the element, and results in more abrupt and uniform switching of the element. This behavior is well corroborated by the switching sequences observed for $f = 1.8$ GHz and $f = 2.5$ GHz, which in turn suggests that the range of $1.8 - 2.5$ GHz falls within the natural resonance frequency region of the element. The majority of the element has completed magnetization reversal at both $f = 1.8$ GHz and $f = 2.5$ GHz by $t = 1.61$ ns, except for the tapered regions at the edges of the element. However, the propagation of the reversed domains to the edges of the element and complete magnetization reversal occur over differing time scales for the two excitation frequencies. For example, at $f = 1.8$ GHz complete switching is achieved within 2.93 ns, while at $f = 2.5$ GHz the switching is complete only by 4.42 ns. As the demagnetizing field profiles at the element edges are the same in both cases, this observation may be attributed to the different precession angles at the two microwave excitation frequencies. As a result, the component of the applied microwave field along the hard axis of the magnetization in these regions is different for each frequency.

Another local anomaly in the switching characteristic is observed at $f = 2.0$ GHz, where the switching sequence by domain nucleation at the foci and subsequent propagation is similar to that of the element in Figs. 1(c-h). In fact, the domain evolution up till $t = 1.77$ ns at $f = 2.0$ GHz is similar to the non-successful switching attempt at $f = 1.5$ GHz. Beyond $t = 1.77$ ns, the element undergoes domain annihilation and gradual relaxation to its initial state at $f = 1.5$ GHz, but complete magnetization reversal happens within 4 ns at $f = 2$ GHz. Thus, the success of magnetization switching for fixed values of applied microwave and static fields depends on the frequency of the microwave excitation. Similar dependence of the switching field on the microwave frequency has been reported in micromagnetic simulations[25] on elements of dimensions $1024 \times 128 \times 20$ nm$^3$. To confirm these similarities in switching behavior, we performed micromagnetic simulations (see Methods) to estimate the effective resonance strength across microwave excitation frequencies. As the precession angle varies with microwave excitation frequency, the standard deviation ($\Delta M_{eff}$) of the simulated magnetization is assumed to be a measure of the effective resonance strength. Figure 3(b) shows the variation of $\Delta M_{eff}$ as a function of the applied microwave excitation frequency. Analogous to the experimental data, we observe larger resonances at frequencies in the range of $1.8 - 2.5$ GHz as compared to that at $f = 1.5$ GHz. We find that spatially-varying demagnetizing field profiles and edge pinning field profiles can lead to a range of precessional frequencies, and in turn cause spin wave generation.

**Discussion**

The initiation of the switching has been previously attributed in literature to the formation of instability regions as mentioned earlier and the origin of these instabilities comes from spin wave generation. However, it is unclear to what extent spin wave generation assists in nucleation and magnetization reversal. To substantiate the influence of high-frequency spin waves in the switching process, Fourier analysis of the simulated data has been carried out in the time domain to identify the dominant frequencies. Figures 4(a-f) show the simulated magnetization profiles of the element at different stages in the reversal process, in a color scale similar to that of the experimental time-resolved images in Figs. 1(c-h). It is important to note that while the reversal stages in experiment and simulation match well, the overall process occurs at a faster rate in simulations due to the lack of any material and physical deformities, such as patterning defects. To identify the dominant frequency modes in the bulk of the magnetic element, the Fourier analysis was performed on the points along the easy axis, as indicated in the inset of Fig. 4(g). The analyzed locations pass through the magnetically-soft central regions of the element where the in-plane demagnetizing field is the lowest. Therefore, any disturbance or dynamics would be maximized along this line. Figure 4(g) shows the FFT for the $m_x$ data for a total simulation time of 6 ns. The regions between points $75 - 125$ and $225 - 275$ represent the foci of the elliptically-shaped element, where the internal field is at a minimum and the observed FFT spectra is at its highest intensity. This spectra exactly matches the domain nucleation and switching behavior experimentally seen in the $4 \times 0.4$ $\mu m^2$ element shown in Fig. 4(h) at $t = 1.44$ ns. Clear correlation can be seen between the yellow and red dotted regions in Figs. 4(g,h). Thus, the high intensity peaks in the simulated data at a frequency of ~5.6 GHz should correspond to the local resonance frequency within these domains, as reported earlier[8,26]. This result confirms that spin waves may be a critical factor in a microwave assisted magnetization switching process.

In conclusion, we have directly imaged the magnetization switching dynamics under the influence of a microwave-assisted field pulse in patterned elements for the first time with sub-nanosecond temporal resolution. Excitation with a varying frequency of the microwave pulse reveals the natural resonance frequency range of the element, and confirms the influence of a microwave assist field in magnetization reversal. The Fourier analysis in the time domain reveals the presence of higher local resonant frequencies at the foci of the elliptical element, and matches the switching sequence observed in experimental results. This, in turn, shows that magnetization reversal depends on the spatial variation of the demagnetizing field.

**Methods**

**Sample preparation details**. The Py elements studied in this work were prepared by standard electron beam lithography, sputtering, and liftoff processes. Elliptically-shaped elements of dimensions $4 \times 0.4 \ \mu m^2$ and $6 \times 0.7 \ \mu m^2$ were patterned using a 200 nm-thick Polymethyl methacrylate (PMMA) resist mask. A film stack of Ta (2 nm)/$Ni_{78}Fe_{22}$ (Py, 30 nm) was then deposited in a sputtering chamber with a base chamber pressure of $2 \times 10^{-9}$ Torr. Following lift-off, conventional linear striplines were defined by photolithography using a 2 $\mu m$-thick Sumitomo PFI-D81B8 resist. The Ta/Cu pads were sputter-deposited followed by lift-off processing.

**Time-resolved scanning transmission X-ray microscopy**. The time-resolved imaging data were obtained by magnetic soft X-ray transmission microscopy at the MAXYMUS microscope (Beam-line UE46 PGM-2 at BESSY II synchrotron in Berlin, Germany), maintained by the MPI for Intelligent Systems. X-ray energy was tuned to the Ni $L_3$ absorption edge (852.7 eV), where the XMCD effect yields the largest magnetic contrast for Py. The sample was tilted by 60° with respect to the X-ray beam to measure the $m_x$ and $m_z$ components of the magnetization with a spatial resolution of 25 nm. A fast digital pulse generator and a signal generator in combination with a frequency mixer were used to inject rectangular pulses (pulse of 2 ns duration, and rising & falling times of ~ 300 ps), superimposed with sine wave burst pulses (pulse width of 4 ns) through a strip line to generate the in-plane magnetic field burst pulses. The field strengths are calculated from the currents and have an estimated systematic error of ~ 10 %. In order to increase the signal-to-noise ratio and yield reasonable dynamic contrast, a stroboscopic pump-and-probe technique that averages several million excitations per imaged pixel was used. We set the repetition rate of the excitation to 12 MHz and accumulated the information of the magnetic state every 166 ps, thereby providing good temporal resolution during the reversal process. The multi bunch mode of the synchrotron was synced with the experiment to gain the complete temporal information per measured pixel simultaneously, before moving to the next pixel and so on, thus mapping the temporospatial magnetic state of the sample in one single scan. Beam-time periods at synchrotron facilities are very limited, and are also known to present significant challenges in setting up an initial, successful proof-of-concept experiment. As a result, the switching experiments that we could study were constrained. For these reasons, we only present the data of MAS at two angles of rotation (30° and 45° to the $-x$ axis) of the ferromagnetic element.

**Micromagnetic simulations and Fourier analysis.** The micromagnetic simulations were carried out with the object oriented micromagnetic framework (OOMMF, OOMMF User's Guide, Version 1.0, Interagency Report NISTIR 6376 (National Institute of Standards and Technology, 1999)). We performed the simulations at cell sizes of $(5 \times 5 \times 30)$ nm$^3$ and $(2 \times 2 \times 30)$ nm$^3$, and similar results were obtained in both cases. The exchange interaction length in Py is known to be (~5 nm). Standard material parameters for Py are used in the simulations such as the saturation magnetization $M_s = 860 \times 10^3$ A/m, exchange stiffness $A = 1.3 \times 10^{-11}$ J/m, zero magnetocrystalline anisotropy, and a damping constant $\alpha = 0.01$.

# References

1. Devolder, T., Chappert, C., Katine, J. A., Carey, M. J. & Ito, K. Distribution of the magnetization reversal duration in subnanosecond spin-transfer switching. *Phys. Rev. B* **75**, 064402 (2007).

2. Hui, Z. *et al.* Sub-200 ps spin transfer torque switching in in-plane magnetic tunnel junctions with interface perpendicular anisotropy. *J Phys. D. Appl. Phys.* **45**, 025001 (2012).

3. Liu, H. *et al.* Ultrafast switching in magnetic tunnel junction based orthogonal spin transfer devices. *Appl. Phys. Lett.* **97**, 242510 (2010).

4. Choi, B. C., Belov, M., Hiebert, W. K., Ballentine, G. E. & Freeman, M. R. Ultrafast Magnetization Reversal Dynamics Investigated by Time Domain Imaging. *Phys. Rev. Lett.* **86**, 728-731 (2001).

5. Gerrits, T., van den Berg, H. A. M., Hohlfeld, J., Bar, L. & Rasing, T. Ultrafast precessional magnetization reversal by picosecond magnetic field pulse shaping. *Nature* **418**, 509-512 (2002).

6. Kovalenko, O., Pezeril, T. & Temnov, V. V. New Concept for Magnetization Switching by Ultrafast Acoustic Pulses. *Phys. Rev. Lett.* **110**, 266602 (2013).

7. Paul, P. H. *et al.* Ultra-fast ballistic magnetization reversal triggered by a single magnetic field pulse. *J Phys. D. Appl. Phys.* **42**, 245007 (2009).

8. Stoll, H. *et al.* High-resolution imaging of fast magnetization dynamics in magnetic nanostructures. *Appl. Phys. Lett.* **84**, 3328-3330 (2004).

9. Thirion, C., Wernsdorfer, W. & Mailly, D. Switching of magnetization by nonlinear resonance studied in single nanoparticles. *Nature Mater.* **2**, 524-527 (2003).

10. Han, X. *et al.* Magnetic Instability Regions in Patterned Structures: Influence of Element Shape on Magnetization Reversal Dynamics. *Phys. Rev. Lett.* **98**, 147202 (2007).

11. Grimsditch, M. *et al.* Magnetic stability of nano-particles - The role of dipolar instability pockets. *Europhys. Lett.* **54**, 813-819 (2001).

12. Yanes, R. *et al.* Modeling of microwave-assisted switching in micron-sized magnetic ellipsoids. *Phys. Rev. B* **79**, 224427 (2009).

13. Hiebert, W. K., Ballentine, G. E., Lagae, L., Hunt, R. W. & Freeman, M. R. Ultrafast imaging of incoherent rotation magnetic switching with experimental and numerical micromagnetic dynamics. *J. Appl. Phys.* **92**, 392-396 (2002).

14. Bauer, M., Lopusnik, R., Fassbender, J. & Hillebrands, B. Suppression of magnetic-field pulse-induced magnetization precession by pulse tailoring. *Appl. Phys. Lett.* **76**, 2758-2760 (2000).

15. Gadbois, J., Zhu, J. G., Vavra, W. & Hurst, A. The effect of end and edge shape on the performance of pseudo-spin valve memories. *IEEE Trans. Magn.* **34**, 1066-1068 (1998).

16. Zhang, W. L. *et al.* Magnetization reversal simulation of diamond-shaped NiFe nanofilm elements. *IEEE Trans. Magn.* **41**, 4390-4393 (2005).

17. Nozaki, Y., Ishida, N., Soeno, Y. & Sekiguchi, K. Room temperature microwave-assisted recording on 500-Gbpsi-class perpendicular medium. *J. Appl. Phys.* **112**, 083912 (2012).

18. Nozaki, Y., Narita, N., Tanaka, T. & Matsuyama, K. Microwave-assisted magnetization reversal in a Co/Pd multilayer with perpendicular magnetic anisotropy. *Appl. Phys. Lett.* **95**, 082505 (2009).

19. Nozaki, Y. *et al.* Microwave-assisted magnetization reversal in 0.36-μm-wide Permalloy wires. *Appl. Phys. Lett.* **91**, 122505 (2007).

20. Zhu, J. G. & Wang, Y. Microwave Assisted Magnetic Recording Utilizing Perpendicular Spin Torque Oscillator With Switchable Perpendicular Electrodes. *IEEE Trans. Magn.* **46**, 751-757 (2010).

21. Zhu, J. G., Zhu, X. & Tang, Y. Microwave Assisted Magnetic Recording. *IEEE Trans. Magn.* **44**, 125-131 (2008).

22. Woltersdorf, G. & Back, C. Microwave Assisted Switching of Single Domain $Ni_{80}Fe_{20}$ Elements. *Phys. Rev. Lett.* **99**, 227207 (2007).

23. Martín Pimentel, P., Leven, B., Hillebrands, B. & Grimm, H. Kerr microscopy studies of microwave assisted switching. *J. Appl. Phys.* **102**, 063913 (2007).
24. Schütz, G. *et al.* Absorption of circularly polarized X rays in iron. *Phys. Rev. Lett.* **58**, 737-740 (1987).
25. Laval, M., Bonnefois, J. J., Bobo, J. F., Issac, F. & Boust, F. Microwave-assisted switching of NiFe magnetic microstructures. *J. Appl. Phys.* **105**, 073912 (2009).
26. Park, J. P., Eames, P., Engebretson, D. M., Berezovsky, J. & Crowell, P. A. Imaging of spin dynamics in closure domain and vortex structures. *Phys. Rev. B* **67**, 020403 (2003).

**Acknowledgments**

This work is supported by the Singapore Ministry of Education Academic Research Fund Tier 1 (R-263-000-A46-112).

**Author contributions**

S.R. and J.R. fabricated devices. S.R., J.R., A.B., and M-A.M. carried out x-ray measurements, and M.W. and M.K. helped experiments. S.R., J.R., A.B., C.B., and H.Y. analyzed the experiments. S.R. and H.Y. wrote the paper. All authors discussed the results and commented the paper. H.Y. supervised the project.

**Additional information**

**Supplementary information** accompanies this paper at http://www.nature.com/ scientificreports

**Figure captions**

**Figure 1| Sample schematic and experimental details of the STXM-XMCD measurement.** (a) The sample consists of elliptically-shaped magnetic Py elements (red) below a current-carrying stripline. The sample is aligned at 60° with respect to the X-ray beam. The current pulse $I(t)$ through the stripline consists of a 2 ns square pulse superimposed on a 4 ns microwave signal (at 1.8 GHz), as shown in (b). The inset in (b) shows a STXM-XMCD differential image of the Py element, along with the relative directions of the pulse and microwave magnetic fields. (c–h) Time-resolved images of the reversal process in a $4 \times 0.4$ $\mu m^2$ Py element.

**Figure 2| Switching experiments under the influence of a single excitation current pulse.** (a–f) Partial switching in a $4 \times 0.4$ $\mu m^2$ magnetic element with a square pulsed current and no microwave current. (g–i) Weak excitations imaged in the element for only a microwave current excitation at 1.8 GHz without the square pulse, as indicated by the red arrows.

**Figure 3| Switching experiments under the influence of a varying microwave excitation frequency.** (a) Time-resolved images of the reversal process in a $6 \times 0.7$ $\mu m^2$ Py element for different excitation frequencies of the microwave current pulse. (b) Measure of the effective resonance strength ($\Delta M_{eff}$) as a function of applied microwave excitation frequency ($f_{RF}$), from simulated magnetization $m_x$.

**Figure 4| Simulated magnetization reversal images and Fast Fourier Transform (FFT) in the frequency domain.** (a-f) Snapshots of the simulated magnetization in an elliptical element undergoing microwave assisted switching. The color scale used is white-gray-black corresponding to the experimental contrast. (g) FFT in frequency domain performed on the simulated data of the magnetization $m_x$ at points along the easy axis of the element indicated by

red line in the inset. (h) Time-resolved STXM-XMCD differential image of the $4 \times 0.4$ $\mu m^2$ magnetic element at $t = 1.44$ ns, with partially switched domains at the foci of the elliptically-shaped element. The dotted regions in (g) and (h) show the correlation between simulated and experimental data in the frequency and spatial domains, respectively.

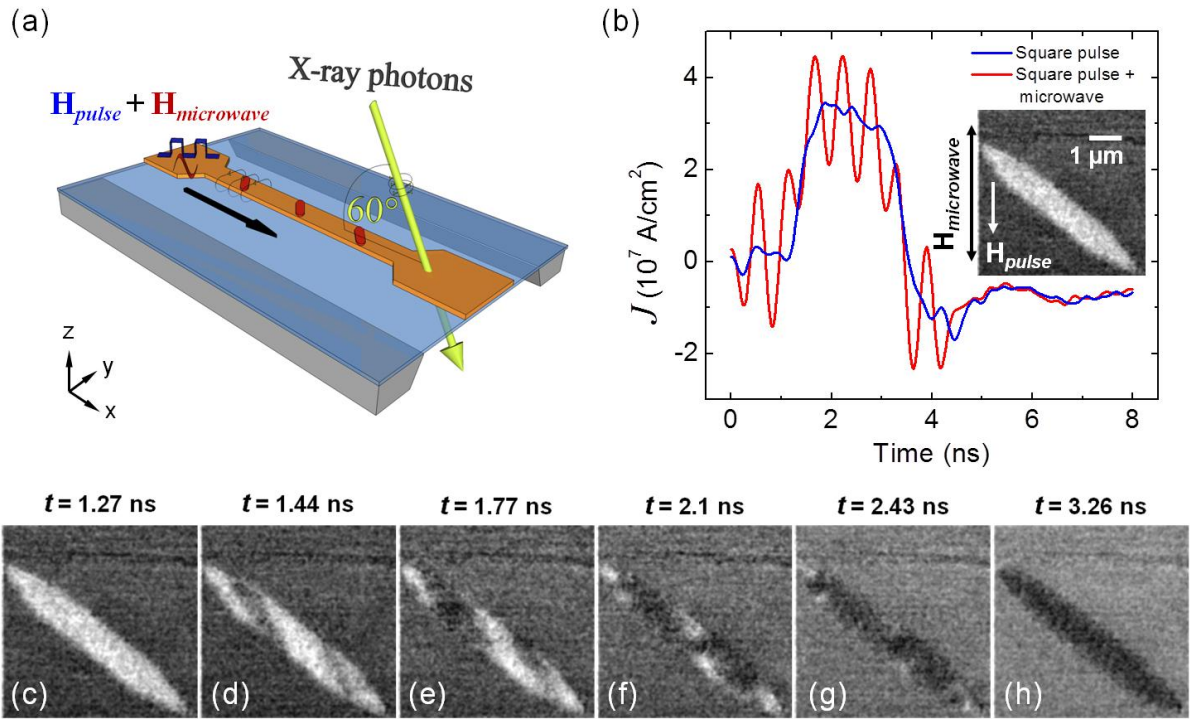

(a) **H**$_{pulse}$ + **H**$_{microwave}$    X-ray photons    60°

z
y
x

(b)

$J$ ($10^7$ A/cm$^2$)

— Square pulse
— Square pulse + microwave

Time (ns)

1 µm

**H**$_{microwave}$
**H**$_{pulse}$

| $t$ = 1.27 ns | $t$ = 1.44 ns | $t$ = 1.77 ns | $t$ = 2.1 ns | $t$ = 2.43 ns | $t$ = 3.26 ns |

(c)    (d)    (e)    (f)    (g)    (h)

Fig. 1

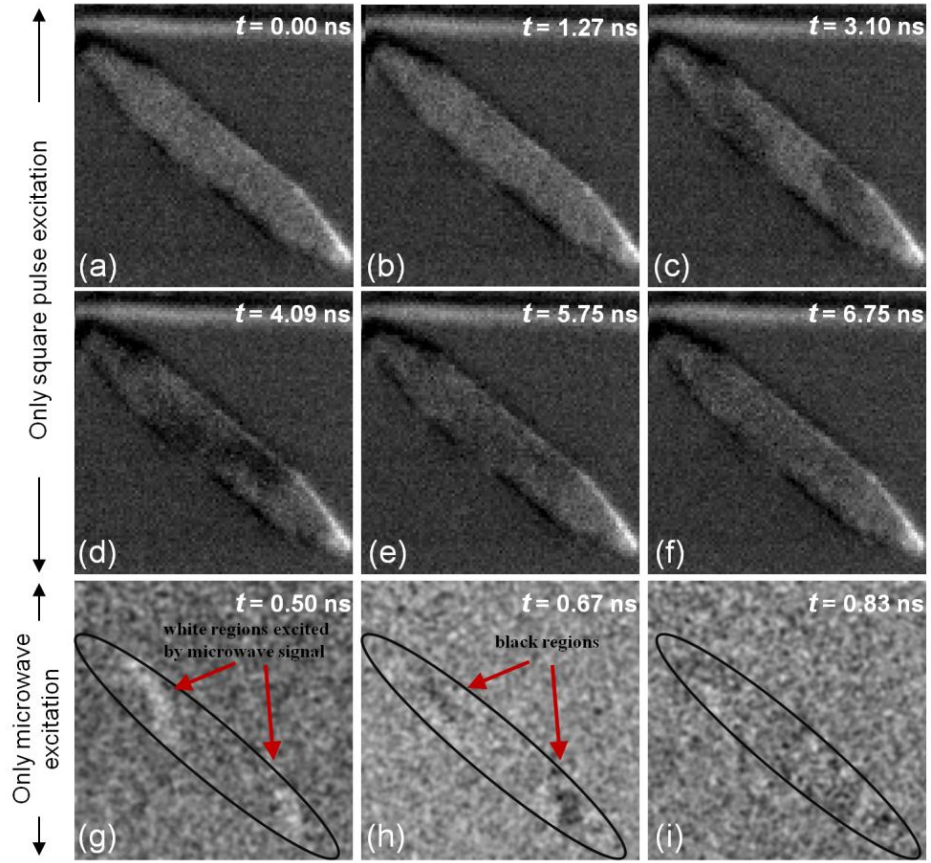

Fig. 2

(a)

| $f_{RF}$ = 1.5 GHz | $f_{RF}$ = 1.8 GHz | $f_{RF}$ = 2.0 GHz | $f_{RF}$ = 2.5 GHz |

Time elapsed

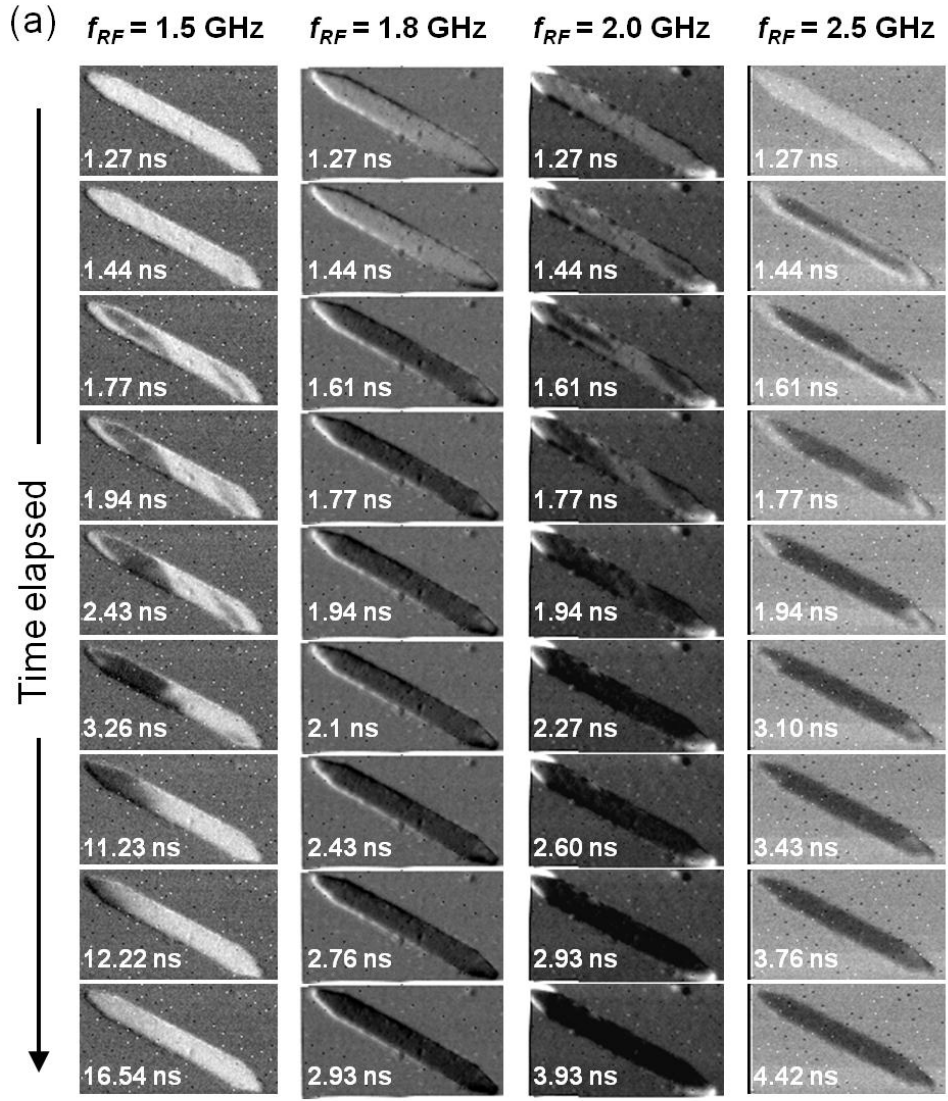

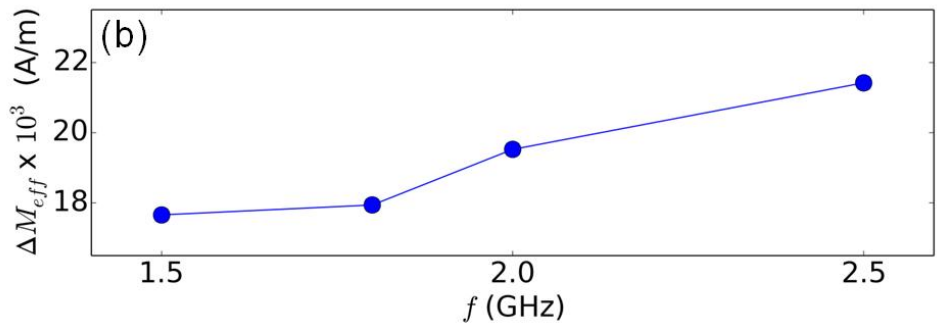

Fig. 3

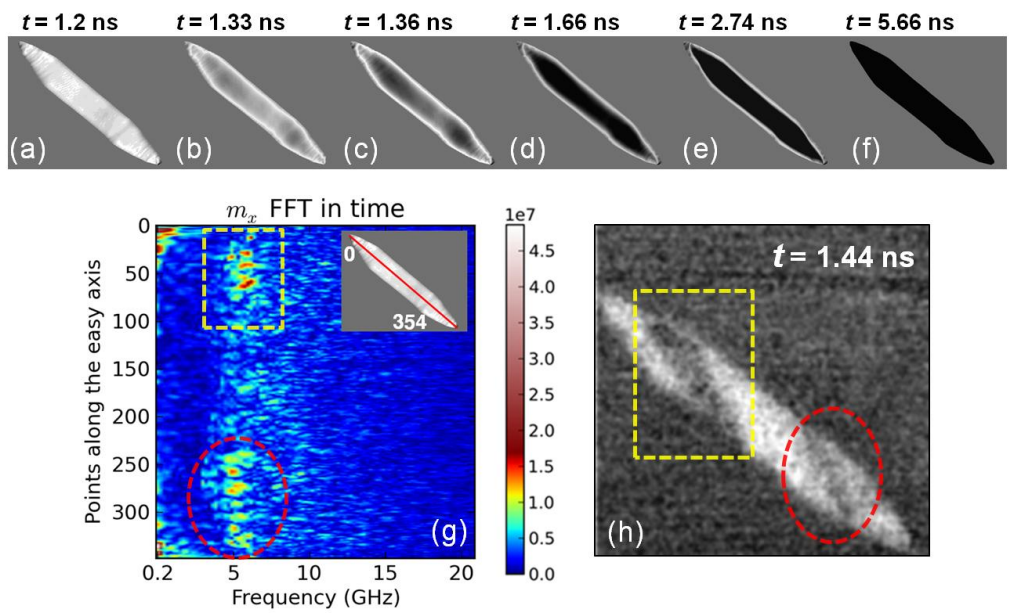

*t* = 1.2 ns  *t* = 1.33 ns  *t* = 1.36 ns  *t* = 1.66 ns  *t* = 2.74 ns  *t* = 5.66 ns

(a) (b) (c) (d) (e) (f)

$m_x$ FFT in time

(g) (h) *t* = 1.44 ns

Fig. 4